\documentclass[pdflatex,sn-mathphys-num]{sn-jnl}

\usepackage{appendix}
\usepackage{graphicx}%
\usepackage{multirow}%
\usepackage{amsmath,amssymb,amsfonts}%
\usepackage{amsthm}%
\usepackage{mathrsfs}%
\usepackage[title]{appendix}%
\usepackage{xcolor}%
\usepackage{textcomp}%
\usepackage{manyfoot}%
\usepackage{booktabs}%
\usepackage{algorithm}%
\usepackage{algorithmicx}%
\usepackage{algpseudocode}%
\usepackage{listings}%
\usepackage{comment}
\usepackage{placeins}
\newcommand{\mo}{\textcolor{black}} 

\theoremstyle{thmstyleone}%
\theoremstyle{thmstyletwo}%

\theoremstyle{thmstylethree}%

\begin{document}

\title[Article Title]{Resonant interactions in the $\alpha$-FPUT lattice with site-dependent coefficients
}

\author[1,2]{\fnm{Lorenzo} \sur{Migliorelli}}

\author*[1,2]{\fnm{Giovanni } \sur{Dematteis}}\email{giovannidematteis@gmail.com }

\author[3]{\fnm{Sergio} \sur{Chibbaro}}

\author[1,2]{\fnm{Miguel} \sur{ Onorato}}

\affil[1]{\orgdiv{Dipartimento di Fisica}, \orgname{Universit\'a degli Studi di Torino}, \orgaddress{\street{Via P. Giuria 1}, \postcode{10125} \state{Torino}, \country{Italy}}}

\affil[2]{\orgdiv{Istituto Nazionale di Fisica Nucleare, INFN}, \orgname{Sezione di Torino}, \orgaddress{\street{Via P. Giuria 1}, \postcode{10125} \state{Torino}, \country{Italy}}}

\affil[3]{\orgdiv{Universit\'e Paris-Saclay}, \orgname{LISN, CNRS, INRIA, UMR 9015}, \orgaddress{ \city{Orsay cedex}, \postcode{F-91405  }, \country{France}}}

\abstract{The wave turbulence framework has proven to be an effective tool for analyzing certain features of nonlinear energy transfer in one-dimensional nonlinear chains. In this work, we extend this approach to the $\alpha$-FPUT problem when the spring stiffness $\chi$ and the nonlinear coefficient $\alpha$ are site-dependent. Although three-wave interactions are non-resonant for constant coefficients, their spatial modulation gives rise to a non-trivial resonant manifold.  In this framework, we derive a new kinetic equation that suggests the possibility of substantially faster thermalization than in the constant-coefficient case. The new kinetic equation also includes an additional term that can be associated with the Bragg-scattering mechanism, which promotes the isotropization of the wave-action spectral density function. {\color{black}{We corroborate these results with direct numerical simulations, confirming the radically faster thermalization pathway in the presence of random coefficients.}}}

\keywords{FPUT, Wave Turbulence, Thermalization}



\maketitle

\section{Introduction}\label{sec1.1}
The study of one-dimensional anharmonic chains occupies a central position in nonlinear dynamics and statistical physics~\cite{lieb2013mathematical}. These systems, consisting of coupled particles interacting via nonlinear potentials, provide a fertile ground for investigating fundamental questions about energy transfer, thermalization, and wave dynamics. Their mathematical simplicity belies a rich variety of phenomena, ranging from integrable dynamics to chaotic behavior, and their relevance spans disciplines such as condensed matter physics, nonlinear optics, and even cosmology.

The interest in anharmonic chains dates back to the mid-twentieth century, when physicists began exploring deviations from ideal harmonic behavior in lattice dynamics. The pioneering work of Fermi, Pasta, Ulam, and Tsingou (FPUT) \cite{fermi1955studies} marked a turning point. Their numerical experiments revealed unexpected recurrences in energy distribution among modes, challenging the classical assumption that nonlinear interactions always lead to equipartition of energy at equilibrium. This discovery  stimulated research into nonlinear wave equations, chaos theory, and integrability~\cite{gallavotti2007fermi,campbell2005introduction}.
Beyond the FPUT problem, anharmonic chains have been studied in various contexts, such as heat conduction \cite{lepri2003thermal}, energy localization \cite{flach1998discrete}, and the emergence of solitons \cite{zabusky1965interaction}. These systems are also intimately connected to foundational questions in statistical mechanics, including the mechanisms of thermalization and the role of ergodicity breaking~\cite{castiglione2008chaos,baldovin2025foundations}.

Anharmonic chains are typically described by Hamiltonians that include both quadratic (harmonic) and higher-order (anharmonic) terms. A general form of the Hamiltonian is
\begin{equation}
H = \sum_{i=1}^N \left[ \frac{p_i^2}{2} + V(q_{i+1} - q_i) \right],
\end{equation}
where $q_i$ and $p_i$ represent the displacement and momentum of the $i$-th particle, and $V(r)$, where $r=q_{i+1} - q_i$ is the interparticle potential. For anharmonic systems, $V(r)$ typically includes nonlinear terms, such as:
\begin{equation}
V(r) = \frac{1}{2}r^2 + {\frac{\alpha}{3}} r^3 + {\frac{\beta}{4}} r^4.
\end{equation}

Specific models, such as the FPUT chain, focus on polynomial potentials, while others, like the Toda lattice \cite{toda1967vibration}, use exponential potentials to explore integrable dynamics ~\cite{benettin2013fermi}. These variations allow researchers to probe different aspects of nonlinear interactions and statistical properties.
A key question in the study of anharmonic chains is how energy redistributes among modes and whether the system reaches thermal equilibrium. The equipartition theorem predicts uniform energy distribution in harmonic systems, but anharmonic interactions introduce complexities such as mode coupling and localization~\cite{fucito1982approach,livi1985equipartition,benettin2005time}. 
Although many insights have been gained through numerical simulations and dynamical systems tools~\cite{casetti1997fermi,benettin2013fermi,Benettin2023FPUmodel}, Wave Turbulence (WT) has been found to be relevant for understanding the underlying mechanisms of the different scaling regimes~\cite{onorato2015route,onorato2023wave}.

Generally speaking, Wave Turbulence theory provides a framework for understanding energy transfer in weakly nonlinear systems~\cite{zakharov2025kolmogorov,nazarenko2011wave,newell2011wave,galtier2022physics}. By treating the dynamics statistically, it is possible to derive, at least in a formal way, the so-called wave kinetic equations, see \cite{zakharov2025kolmogorov, hasselmann1962non, nazarenko2011wave, ZakharovFilonenkoOmegaMinusFour}, describing the spectral energy density evolution in time.
Wave Turbulence arises when waves of different wavelengths interact through nonlinear mechanisms, leading to the redistribution of energy across scales. In weakly nonlinear regimes, these interactions can be described statistically using the wave kinetic equation derived from the underlying wave dynamics. The approach is grounded in the assumption that the nonlinear interactions are slow compared to the linear wave propagation, allowing for a perturbative treatment.
Wave Turbulence constitutes a broad and versatile framework that has been successfully applied in a wide range of physical contexts, including gravity waves, Bose--Einstein condensates, nonlinear optics, plasma waves, elastic media, internal waves in stratified fluids, capillary waves, magnetohydrodynamics, gravitational waves, Kelvin waves, inertial waves in rotating fluids, and lattice systems \cite{zakharov2025kolmogorov,nazarenko2011wave,galtier2022physics,picozzi2014optical}. Many geophysical predictions are based on wave kinetic models, and the field continues to attract significant interest from physicists, geophysicists, and mathematicians.
The development of wave kinetic theory dates back to the pioneering works of Peierls \cite{peierls1929} on phonon interactions in crystals and of Hasselmann and  Zakharov  on wave interactions in the ocean \cite{hasselmann1962non,ZakharovFilonenkoOmegaMinusFour}. 
The wave kinetic equation predicts that an irreversible transfer of energy takes place when, at some order in nonlinearity, the resonant condition is satisfied. For example for quadratic nonlinearity in the equation of motion, the resonant condition implies that three wavenumbers satisfy simultaneously the following conditions 
\begin{equation}
\begin{split}
k_1 &= k_2 \pm k_3, \\
\omega_1 &= \omega_2 \pm \omega_3,
\end{split}
\end{equation}
where $\omega_i=\omega(k_i)$ are the frequencies associated with the wavenumbers through the dispersion relation. 
The kinetic equation captures the essential mechanisms governing the transfer of energy and wave action in weakly nonlinear wave systems. As will be discussed in the following sections, it admits, as stationary solution, the thermodynamic equilibrium state given by the Rayleigh-Jeans distribution, which generalizes the equipartition of energy among Fourier modes in systems that conserve not only energy, but also wave action (and, in some cases, momentum). 
While the wave kinetic equation is widely used in the physics literature, its rigorous derivation remains a challenging problem, with only a limited number of results currently available \cite{buckmaster2020kinetic,staffilani2021wave,deng2021derivation,deng2023long,wu2025rigorous}.

After the pioneering work by Pereverzev~\cite{pereverzev2003fermi} and then by Spohn and collaborators, \cite{spohn2006phonon, aoki2006energy,lukkarinen2008anomalous,lukkarinen2016kinetic}, the Wave Turbulence approach has also proven to be a powerful and unifying framework for understanding the long-time dynamics and thermalization time scale of the $\alpha$- and $\beta$-FPUT chains in the weakly nonlinear regime, see \cite{onorato2023wave}. 
 Applied to the FPUT chain, it shows that irreversible energy transfer is driven by  resonances, whose structure critically depends on the type of nonlinearity. In particular, for the $\alpha$-FPUT model, three-wave interactions are non-resonant in one dimension, and thermalization is therefore mediated by higher-order processes ~\cite{onorato2015route}, resulting in very long equipartition times. In contrast, the $\beta$-FPUT model admits, in the large-box limit, nontrivial four-wave resonances, leading to an efficient redistribution of energy \cite{lvov2018double}.  These theoretical predictions are well supported by numerical simulations, both for finite systems and in the thermodynamic limit~\cite{pistone2018universal,zhang2021behaviors,lin2025universality}. 
 Overall, the Wave Turbulence approach has been instrumental in clarifying the original FPUT paradox, showing that thermalization does occur, albeit on time scales determined by the structure and efficiency of resonant interactions.

The aim of the present work is to analyze theoretically a different setup of the FPUT problem.
Namely, we consider the $\alpha$-FPUT case with variable coefficients. From a physical point of view, this means the springs are not all identical and the corresponding harmonic and anharmonic constants depend on the lattice site. 
This case is related with the problem of disordered (random) harmonic chains which have been widely studied in the linear case~\cite{dyson1953dynamics,matsuda1970localization,casher1971heat}, and, more recently, in the presence of nonlinear coupling~\cite{pikovsky2008destruction,mulansky2011strong,mulansky2013energy}.
Here, we propose to deal with the general yet simpler case in which the spring coefficient is weakly 
dependent on the position.  
This has the important advantage, as we shall show, of allowing for a straightforward application of the WT theory. 
It turns out that this problem is `different from that of the standard $\alpha$-FPUT chain. Notably, lower-order resonances arise, leading to a different, and more efficient transfer of energy among modes.
Then, we  build the corresponding kinetic equation, which points to the presence of a linear three-wave term, and a new nonlinear term related to interactions that involve four wavenumbers but only three frequencies, while only pure 4-wave resonances are permitted in the standard $\alpha$-FPUT case. We interpret this kind of interaction as 3-wave+1 resonances, the new ``wave'' being related to the variability of the coupling coefficient in space.
Interestingly, a linear and a nonlinear term appear in the equation. The linear term is shown to be related to Bragg-scattering, whereas the nonlinear term is proportional to the square of the small nonlinearity parameter $\epsilon$. 
Therefore, our theoretical prediction is that the characteristic energy-transfer timescale for the present case should be of the order of $1/\epsilon^{2}$, i.e. much shorter than that of the standard $\alpha$-FPUT case which scales as proportional to $1/\epsilon^{4}$ \cite{onorato2015route,onorato2023wave}.

The paper is organized as follows: In Section \ref{sec:model}, we present the mathematical model.  In Section \ref{sec:resonances}, we show the existence of a resonant manifold. Then, in Section \ref{sec:normalform} we derive a reduced system by eliminating non-resonant interactions through a normal form expansion. This latter system is suitable for a statistical description and the associated wave kinetic equation is presented in Section \ref{sec:wke} (the details of the derivation are presented in the appendix). \mo{Direct numerical simulations verifying the theoretical predictions are presented in Section \ref{sec:num_comp}.} We present a discussion and our conclusions in Section \ref{sec:conclusions}.

\section{The $\alpha$-FPUT with variable coefficients}\label{sec:model}
\mo{We consider the following Hamiltonian
for $N$ equal masses connected by springs:
\begin{equation}
\mathcal{H}=\sum_j\left(\frac{p_j^2}{2m}+\frac{\chi_j}{2}(q_{j+1}-q_j)^2
    +\varepsilon\frac{\alpha_j}{3}(q_{j+1}-q_j)^3\right)
\end{equation}}

The equations of motion of the $\alpha$ -FPUT chain take the following form:
\begin{equation}
m\ddot q_j=
\mo{\chi_j \left(q_{j+1}-q_j\right)+\chi_{j-1}\left(q_{j-1}-q_j\right)+
\epsilon\left[ \alpha_j (q_{j+1}-q_{j})^2 -\alpha_{j-1} (q_{j-1}-q_{j})^2\right]},
\label{eq:alpha}
\end{equation}
where $m$ is the mass of each oscillator, $\chi_j$ and $\alpha_j$ are coefficients characterizing the linear and nonlinear properties of the springs. The variable $q_j$ denotes the displacement with respect to the equilibrium position. Note that both depend on the site $j$. We have introduced $\epsilon$ as a small parameter in front of the nonlinear term. Furthermore, to be in a regime that is dominated by the standard dispersion relation, we assume that the stiffness constant is given by its mean value, $\bar \chi$ plus some space-dependent fluctuations,  $\chi_j'$, that are of order $\epsilon$:
\begin{equation}
\chi_j=\bar \chi(1+\epsilon \chi_j').
\end{equation}
{\color{black}We choose $\chi_j'$ and $\alpha_j$ as a realization from a random distribution that we characterize in Fourier space (therefore periodic in physical space) by prescribed spectral amplitudes and random phases -- see section 6 for a specific example.}
From now on, without loss of generality, we set $m=\bar\chi=1$. \mo{Additionally, to simplify the notation, we will drop the prime and denote the spatial fluctuations $\chi_j'$ simply as $\chi_j$ throughout the rest of the paper.} We impose periodic boundary conditions on a chain of N sites and expand the wave field and coefficients in Fourier series. We introduce the normal variables $a_k$ as 
\begin{equation}
a_k=\frac{1}{\sqrt{2\omega_k}}\left(\omega_k q_k+i p_k\right),
\end{equation}
\begin{equation}
a_{N-k}^*=\frac{1}{\sqrt{2\omega_k}}\left(\omega_k q_k-i p_k\right),
\end{equation}
here $q_k$ is the Fourier amplitude of $q_j$ and $p_k=dq_k/dt$ and
\begin{equation}
\omega_k=2 \sin(\pi k/N), \;\;\; 1\le k\le N-1
\end{equation}
The equations of motion for $a_k$ take the following form:
\begin{equation} \label{eq:dyn_eq_a}
\begin{split}
&i\frac{da_1}{dt}=\omega_1 a_1+\epsilon \sum_{k_2,k_3}\left(F_{1,2}\chi_3 a_2\delta_{1}^{2,3}+F_{1,-2} \chi_3 a_2^*\delta_{1,2}^{3}\right)+
\\+
&\epsilon\sum_{k_2, k_3, k_4}\alpha_4\left( \mo{V_{1,2,3}}a_2 a_3\delta_1^{2,3,4}+2 \mo{V_{1,-2,3}}a_2^*a_3\delta_{1,2}^{3,4}+\mo{V_{1,-2,-3}}a^*_2a_3^* \delta_{1,2,3}^{4}\right)
\end{split}
\end{equation}
where 
\begin{equation}
\mo{F_{1,2}=\frac{\,2
    \sin\!\left(\frac{\pi k_1}{N}\right)\sin\!\left(\frac{\pi k_2}{N}\right)
    e^{-i\frac{\pi(k_1-k_2)}{N}}}{\sqrt{\omega_1\omega_2}}}
\end{equation}

and
\begin{equation}
\mo{V_{1,2,3}=\frac{\,4i
    \sin\!\left(\frac{\pi k_1}{N}\right)\sin\!\left(\frac{\pi k_2}{N}\right)
    \sin\!\left(\frac{\pi k_3}{N}\right)
    e^{-i\frac{\pi(k_1-k_2-k_3)}{N}}}{\sqrt{2}\,\sqrt{\omega_1\omega_2\omega_3}}}.
\end{equation}

\mo{Negative indices indicate that the corresponding wavenumber $k_i$ is replaced with $N-k_i$. Moreover, we use numerical indices as a shorthand for the wavenumbers (e.g., $1, 2, 3$ stand for $k_1, k_2, k_3$). }Note that the Kronecker deltas mean $\delta_{i,j,\ldots}^{l,m,\ldots}$ implies $k_i+k_j+\ldots=k_l+k_m+\ldots$ and should be intended as $\pmod{N}$ because of the periodicity of the Fourier space.

\section{Resonances}\label{sec:resonances}
In the WT framework, a statistically irreversible transfer of energy occurs when the resonant conditions, associated with momentum and energy conservation, are satisfied. As will be shown in the next section, this is supported by the wave kinetic equation which describes the temporal evolution of the wave action spectral density function.

Since the theory is developed in the large-box limit, the wavenumbers are no longer discrete but continuous, i.e.\ $k \in \mathbb{R}$. In this framework, it is straightforward to show that the linear dispersion relation reduces to
\begin{equation}
\omega(k) = 2 \sin\left(\frac{k}{2}\right), \qquad 0 \le k < 2\pi,
\end{equation}
with $k \in \mathbb{R}$ understood modulo $2\pi$.
Note that in ~(\ref{eq:dyn_eq_a}) the second term in the first sum and the last term in the second sum are never resonant, since the sum of two, or three positive frequencies cannot vanish (unless they are all zero). For the other terms,  the resonant conditions associated with the dynamical equations in (\ref{eq:dyn_eq_a}) are the following: 

i)
\begin{equation}
\begin{cases}
k_1 = k_2 + k_3 \pmod{2\pi}, \\
\omega(k_1) = \omega(k_2),
\end{cases}
\end{equation}
which has solution $k_1=\pm k_2$. The only relevant one is $k_1=-k_2$, so that $2k_1=k_3$. This is the well-known Bragg scattering process by which one wave is reflected if its wavelength is twice the typical periodicity scale of the spring coefficient.

ii)
\begin{equation}
\begin{cases}
k_1 = k_2 + k_3 + k_4 \pmod{2\pi}, \\
\omega(k_1) = \omega(k_2) + \omega(k_3),
\end{cases}
\end{equation}

iii)
\begin{equation}
\begin{cases}
k_1 = -k_2 + k_3 + k_4 \pmod{2\pi}, \\
\omega(k_1) = -\omega(k_2) + \omega(k_3).
\end{cases}
\end{equation}

For $k_4 \ne 0$, it is possible to determine the wavenumbers $k_1$ and $k_4$, given $k_2$ and $k_3$, that satisfy the above resonant conditions. The solutions of the two sets of equations are given by
\begin{equation}
k_1 = 2\pi \pm 2 \arcsin\left[\sin\left(\frac{k_2}{2}\right) + \sin\left(\frac{k_3}{2}\right)\right], 
\qquad
k_4 = k_1 - k_2 - k_3 \pmod{2\pi},
\end{equation}
and
\begin{equation}
k_1 = 2\pi \pm 2 \arcsin\left[\sin\left(\frac{k_3}{2}\right) - \sin\left(\frac{k_2}{2}\right)\right], 
\qquad
k_4 \equiv k_1 + k_2 - k_3 \pmod{2\pi}.
\end{equation}

\section{The normal form expansion} \label{sec:normalform}
To remove the non-resonant terms from the dynamical equation, we  perform a near-identity transformation of the form~\cite{nazarenko2011wave}:
\begin{equation}
    a_1=b_1+\epsilon \left(\sum_{k_2,k_3} G_{1,2}\chi_3 b_2^* \delta_{1,2}^{3}+\sum_{k_2,k_3,k_4}A_{1,2,3}b_2^* b_3^* \alpha_4\delta_{1,2,3}^{4}\right)+O(\epsilon^2),
\end{equation}
with $G_{1,2}$ and $A_{1,2,3}$ to be determined. The evolution equation for $b_k$ reads:
\begin{equation}
\begin{split}
& \frac{db_1}{dt}+ i\omega_1 b_1+
i\epsilon \sum_{k_2,k_3} \big[ (\omega_1+\omega_2) G_{1,2}\chi_3 b_2^* \delta_{1,2}^{3} +
 F_{1,2}\chi_3 b_2\delta_{1}^{2,3} + 
F_{1,-2} \chi_3 b_2^*\delta_{1,2}^{3} \big] 
\\
+ & i\epsilon\sum_{k_2, k_3, k_4}\alpha_4\bigg(\mo{V_{1,2,3}}b_2 b_3\delta_1^{2,3,4} + 2 \mo{V_{1,-2,3}}b_2^*b_3\delta_{1,2}^{3,4} +\mo{V_{1,-2,-3}}b^*_2b_3^* \delta_{1,2,3}^{4}+
\\
&(\omega_1+\omega_2+\omega_3) A_{1,2,3}b_2^* b_3^* \delta_{1,2,3}^{4}\bigg)+O(\epsilon^2) = 0.
\end{split}
\end{equation}
If we set
\begin{equation}
    G_{1,2}=-\frac{\mo{F_{1,-2}}}{\omega_1 + \omega_2}
\end{equation}
and 
\begin{equation}
    A_{1,2,3}=\mo{-\frac{V_{1,-2,-3}}{\omega_1 + \omega_2+\omega_3}},
\end{equation}
then the evolution equation takes the following form:
\begin{equation}\label{starting}
\begin{split}
& i\frac{db_1}{dt}= \omega_1 b_1 +  \epsilon \sum_{k_2,k_3}   F_{1,2}\chi_3 b_2\delta_{1}^{2,3} +
\\
+ & \epsilon\sum_{k_2, k_3, k_4}\alpha_4\left( \mo{V_{1,2,3}}b_2 b_3\delta_1^{2,3,4} + 2 \mo{V_{1,-2,3}}b_2^*b_3\delta_{1,2}^{3,4}\right)+O(\epsilon^2).
\end{split}
\end{equation}
The above equation is the starting point for a statistical analysis and for the derivation of the wave kinetic equation.

\section{The three-wave+1 kinetic equation}\label{sec:wke}

The formal derivation of the wave kinetic equation follows a well-established procedure. Although the starting equation differs slightly from the standard three- or four-wave cases, the overall methodology remains essentially unchanged (see \cite{zakharov2025kolmogorov}). We do not attempt here to provide a fully rigorous mathematical derivation, which may not be possible. 
For completeness, the formal calculation is presented in Appendix \ref{app:derivation}. In the following, we focus instead on the underlying physical assumptions and present the main result: a modified wave kinetic equation that incorporates both a collision term and the Bragg-scattering mechanism.

 As is customary in wave-turbulence theory, we look for a statistical description, i.e., we consider a wave field where amplitudes and phases are random and the main observable is the wave action spectral density function $\langle |b_k|^2\rangle$, where $\langle ...\rangle$ represents the expectation value with respect to some initial data characterized by random phases and amplitudes, \cite{nazarenko2011wave}. Because of the nonlinearity, it turns out that the spectrum depends on a third-order correlator, which in turn depends on a fourth-order correlator. The presence of the small parameter $\epsilon$ allows for a natural closure of the infinite hierarchy of equations. 
After using the Wick decomposition rule, it turns out (see Appendix \ref{app:derivation}) that the evolution of the spectrum takes place on a time scale of order $1/\epsilon^2$.

In the large-box limit ($N\rightarrow{\infty}$, or equivalently $L\rightarrow{\infty}$ by choosing unitary lattice spacing $\Delta x=1$ and thus $L=N\Delta x=N$), then, the sums become integrals and one can define the function $n(k,t)=L\langle |b_k|^2\rangle/(2\pi)\simeq L\langle |a_k|^2\rangle/(2\pi)$. In the long-time limit, the equation becomes:

\begin{equation}
\begin{split}\label{eq:wavekin}
\frac{\partial n_1 }{\partial t}&=2\pi \epsilon^2\Bigg[\int_0^{2\pi }|F_{1,2}|^2 | \chi_3|^2 
(n_2-n_1) \delta(\Delta \omega_{1}^{2})\delta_{1}^{2,3}dk_2dk_3
\\
&\quad
+ 2 \int_0^{2\pi}  \left( \mathcal{R}^1_{2,3,4}-\mathcal{R}^2_{1,3,4}-\mathcal{R}^3_{1,2,4}\right)    dk_2dk_3dk_4
\Bigg]
\end{split}
\end{equation}
with
\begin{equation}
  \mathcal{R}^1_{2,3,4}=|\alpha_{4}|^2|V_{1,2,3}|^2 \delta_{1}^{2,3,4}\delta(\Delta \omega_{2,3}^1)(n_2n_3-n_3n_1-n_1n_2)
\end{equation}
where the deltas now denote Dirac delta functions.

\subsection{The linear case: the Bragg-scattering effect}\label{sec:5.1}
The role of the linear term on the right-hand side (r.h.s.) of Eq.~\eqref{eq:wavekin} is to symmetrize the spectrum through a Bragg scattering process. Indeed, if one neglects the nonlinearity, the equation can be solved analytically. 
Since $\omega(k)$ depends only on $|k|$, it follows that $k_2 = \pm k_1$. The only physically relevant solution is obtained when the resulting wave is reflected, $k_1=-k_2$, which happens under the condition that $2k_1 = k_3$, i.e. the wavelength of the incoming (and reflected) wave is twice the wavelength of the coefficient modulation.  The two deltas can be integrated out to obtain:
\begin{equation}
    \frac{\partial n_k}{dt} = 2\pi \mo{|F_{k,-k}|^2} \frac{|\chi_{2k}|^2}{|d \omega_k/d k|} (n_{-k}-n_k).
\end{equation}
For the initial condition $n_k(0)$, it is straightforward to show that the solution is:
\begin{equation}
n_k(t)=\frac{1}{2}[n_k(0)+n_{-k}(0)]+\mo{\frac{1}{2}}[n_k(0)-n_{-k}(0)]e^{-\mo{4\pi  |F_{k,-k}|^2} \frac{|\chi_{2k}|^2}{|d \omega_k/d k|}t},
\end{equation}
 Any initial condition that is not symmetric will eventually evolve toward a symmetric state. In this symmetric state, the effect of the Bragg contribution vanishes altogether. As a consequence, in the long-time limit the first term can be neglected, leaving only the three-wave+1 interaction term.
 
 \subsection{Conserved quantities and the equipartition of energy}
We are now concerned with the role played by the nonlinear term in Eq.~\eqref{eq:wavekin} regarding relaxation at long times. 
The standard way to study thermodynamic equilibrium goes through the collision invariants of the system, defined as those functions $\gamma_k$ 
such that $\int_0^{2\pi}dk \gamma_k I_k=0$, where $I_k$ is the r.h.s. of Eq.~\eqref{eq:wavekin}, also known as the collision integral. Clearly, the existence of a collision invariant $\gamma_k$ then implies a globally conserved quantity $C=\int_0^{2\pi}dk \gamma_kn_k$, since $dC/dt = \int_0^{2\pi}dk \gamma_k\partial n_k/\partial t=\int_0^{2\pi}dk \gamma_k I_k=0$, by definition of $\gamma_k$.

The only collision invariant of Eq.~\eqref{eq:wavekin} is the linear frequency $\omega_k$, which corresponds to conservation of total linear wave energy $E=\int_0^{2\pi}\omega_kn_k{dk}$. To show this, let us start by defining

\begin{align}
\mathcal{I} &:= \int_0^{2\pi} dk_1 \omega_1I_1 = \mathcal{I}_{\rm L} + \mathcal{I}_{\rm NL}\,,\quad\text{where}\nonumber\\
\mathcal{I}_{\rm L} &:=2\pi\int_0^{2\pi}dk_1dk_2dk_3\; \omega_1 |F_{1,2}|^2|\chi_3|^2
(n_2-n_1)\delta(\Delta\omega_1^2)\delta_1^{2,3}\,,\nonumber\\
\mathcal{I}_{\rm NL}&:=4\pi\int_0^{2\pi}dk_1dk_2dk_3\; \omega_1(\mathcal{R}^1_{2,3,4}-\mathcal{R}^2_{1,3,4}-\mathcal{R}^3_{1,2,4}) \,.\label{eq:27}
\end{align}
We then show separately that both $\mathcal{I}_{\rm L}$ and $\mathcal{I}_{\rm NL}$ vanish. For the former, we can rewrite it as
\begin{align}
    \mathcal{I}_{\rm L}& = \frac12 [\mathcal{I}_{\rm L}+\mathcal{I}_{\rm L}(1\leftrightarrow2)]\nonumber\\
    &= \pi \int_0^{2\pi}dk (\omega_k-\omega_{-k})|F_{k,-k}|^2\frac{|\chi_{2k}|^2}{|d\omega_k/dk|}(n_{-k}-n_{k})=0\,,\label{eq:28}
\end{align}
where $1\leftrightarrow2$ indicates interchange of dummy indices 1 and 2 inside the integral. Moreover, we have integrated over variables $k_2$ and $k_3$ eliminating the two delta functions and considering that only the Bragg-scattering resonance contributes to the integral, as shown in section~\ref{sec:5.1}. The integrand is identically zero because of the parity of the dispersion relation.
Regarding $\mathcal{I}_{\rm NL}$, we rewrite it as
\begin{align}
    \mathcal{I}_{\rm NL} &= \frac13[\mathcal{I}_{\rm NL}+\mathcal{I}_{\rm NL}(1\leftrightarrow2)+\mathcal{I}_{\rm NL}(1\leftrightarrow3)] \nonumber\\
    &= \frac{4\pi}{3}\int_0^{2\pi}dk_1dk_2dk_3\; \big[\omega_1 (\mathcal{R}^1_{234}-\mathcal{R}^2_{134}-\mathcal{R}^3_{124})\nonumber\\
    &\qquad\qquad+\omega_2 (\mathcal{R}^2_{134}-\mathcal{R}^1_{234}-\mathcal{R}^3_{124})+\omega_3 (\mathcal{R}^3_{124}-\mathcal{R}^2_{134}-\mathcal{R}^1_{234})\big]\nonumber\\
    &=\frac{4\pi}{3}\int_0^{2\pi}dk_1dk_2dk_3\; \big[\mathcal{R}^1_{234}(\omega_1-\omega_2-\omega_3)\nonumber\\
    &\qquad\qquad +\mathcal{R}^2_{134}(-\omega_1+\omega_2-\omega_3)+\mathcal{R}^3_{124}(-\omega_1-\omega_2+\omega_3)\big]=0\,.\label{eq:29}
\end{align}
For each of the three contributions to the integrand, the factor containing the frequencies is identical to the argument of the frequency delta function. 
We note that momentum $P=\int_0^{2\pi}kn_k dk$ is not preserved in lattices for which the wavenumber delta function is satisfied modulo $2\pi$ (Umklapp resonances); moreover, the presence of an extra wavenumber associated with the non-constant coefficients prevents momentum from being conserved. Wave action  $N=\int_0^{2\pi} n_k dk$ itself is not preserved by three-wave interactions.

We are ready to define the entropy of the system as
\begin{equation}
    s(t) = \int_0^{2\pi}dk\;\log n_k\,.
\end{equation}
The following {\it $H-$theorem} holds:

If $n_k$ is a solution of the wave kinetic equation, then the entropy grows monotonically over time,
    \begin{equation}
        \frac{ds}{dt} \ge0\,,
    \end{equation}
    and it tends to its maximum value at the thermodynamic equilibrium, where the spectral distribution follows the Rayleigh-Jeans distribution corresponding to energy equipartition:
    \begin{equation}
        \frac{d s}{dt} \stackrel{t\to \infty}{\to}0\,,\quad\text{with} \quad n_k^{\rm(eq)} = \frac{k_BT}{2\pi\omega_k}.\label{eq:RJ}
    \end{equation}

The proof starts by rewriting the time evolution of the entropy functional in terms of the collision integral:
\begin{equation}
    \frac{d s}{dt}=\int_0^{2\pi}dk_1 \frac{1}{n_1}\frac{\partial n_1}{\partial t}=\int_0^{2\pi}dk_1 \frac{I_1}{n_1}\,.\label{eq:H-theorem}
\end{equation}
We note the close similarity to Eq.~\eqref{eq:27}, in which now $\omega_1$ has been replaced by $n_1^{-1}$. We can proceed to rewrite the linear and nonlinear parts of the evolution in the same vein as Eqs.~\eqref{eq:28} and~\eqref{eq:29}. In the integrand of the linear part there explicitly appears a factor \mo{$(n_k-n_{-k})^2/n_kn_{-k}\ge0$}, and all other terms are positive. In the integrand of the nonlinear part containing $\mathcal{R}^1_{234}$, a factor $n_1n_2n_3(n_1^{-1}-n_2^{-1}-n_3^{-1})^2\ge0$ appears and all other terms are positive, with an analogous result for the two terms $\mathcal{R}^2_{134}$ and $\mathcal{R}^3_{124}$. Therefore, $d s/dt\ge0$. When the rate of change of entropy becomes strictly zero, which defines the thermodynamic equilibrium state, the inverse wave-action spectrum $n_k^{-1}$ satisfies the definition of collision invariant and must therefore be proportional to the linear frequency, the only independent collision invariant in the system. The constant of proportionality is fixed by imposing that the total linear energy density $E$ be equal to the temperature $T$ multiplied by the Boltzmann constant $k_B$, yielding the Rayleigh-Jeans distribution~\eqref{eq:RJ}.


\section{Direct numerical Simulations}\label{sec:num_comp}

To corroborate our theoretical findings, we performed direct numerical simulations of the equation of motion~\eqref{eq:alpha} on a chain of $N=1024$ sites with periodic boundary conditions. The system was integrated using a fourth-order symplectic algorithm (PEFRL) with a time step $\Delta t=0.2$. The chain was initialized far from equilibrium: modes up to $k\simeq 160$ are excited with random phases, while the remaining modes carry only a small background energy. The initial state is normalized to have an $\mathcal{O}(1)$ amplitude, ensuring that the parameter $\epsilon$ uniquely controls the strength of the nonlinearity. The statistical results are obtained by averaging over $100$ realizations. For each realization, we drew new random initial phases while keeping the same spatial profile for the coefficients in the case with site-dependent coefficients. 
We initialized the variable coefficients in Fourier space by applying a power-law amplitude spectrum with random phases: $\chi_k = k^{-a}e^{i\phi_k}$ and $\alpha_k = k^{-b}e^{i\psi_k}$. Here, the phases $\phi_k$ and $\psi_k$ are uniformly distributed in the interval $[0, 2\pi]$, with exponents set to $a = 0.75$ and $b=0.5$.

This initialization requires careful consideration. The random phases are necessary to increase the spatial roughness of the coefficients, which is essential to effectively activate the resonances discussed in Section~\ref{sec:resonances}. However, even with a normalization that ensures unit variance, the coefficients may still exhibit large local fluctuations across the $1024$ sites. This poses two main challenges. First, the local nonlinearity may exceed the bounds of the weakly nonlinear regime we aim to probe. Second, and more critically, the Hamiltonian of the $\alpha$-FPUT model is not bounded from below, allowing a spring to ``break'' if the system escapes the cubic potential well. While this breakdown occurs over a finite but long time scale in the standard setting, the introduction of variable coefficients locally alters the height of the potential barrier, which is given by $\Delta V_j = \chi_j^3 / (6\epsilon^2\alpha_j^2)$. To mitigate the risk of breaking the chain, we generated $10000$ independent realizations of the coefficients and selected the specific profile that maximizes the minimum value of $\Delta V_j$ across all sites.

\begin{figure}[h]
    \centering
    \includegraphics[width=0.7\textwidth]{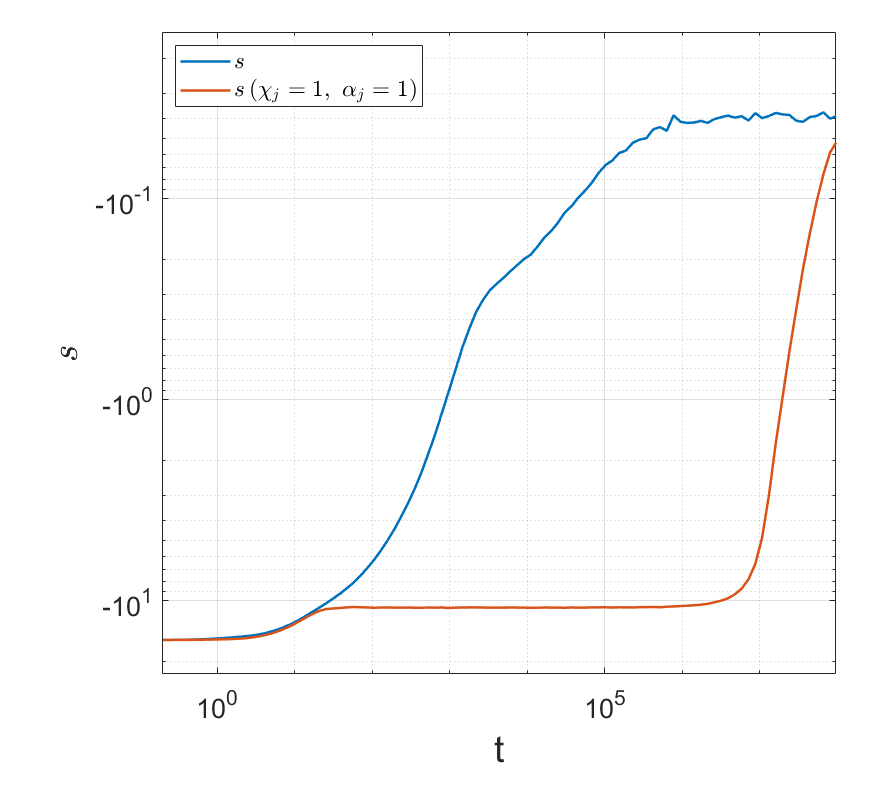}
    \caption{Log-log plot of the entropy $s(t)$. The case with non-constant coefficients (blue) demonstrates a significantly faster approach to thermalization than the baseline case with constant coefficients equal to one (red, $s(\chi_j=1, \alpha_j=1)$).}
    \label{fig:entropy_comparison}
\end{figure}

\begin{figure}[h]
    \centering
    \includegraphics[width=1\textwidth]{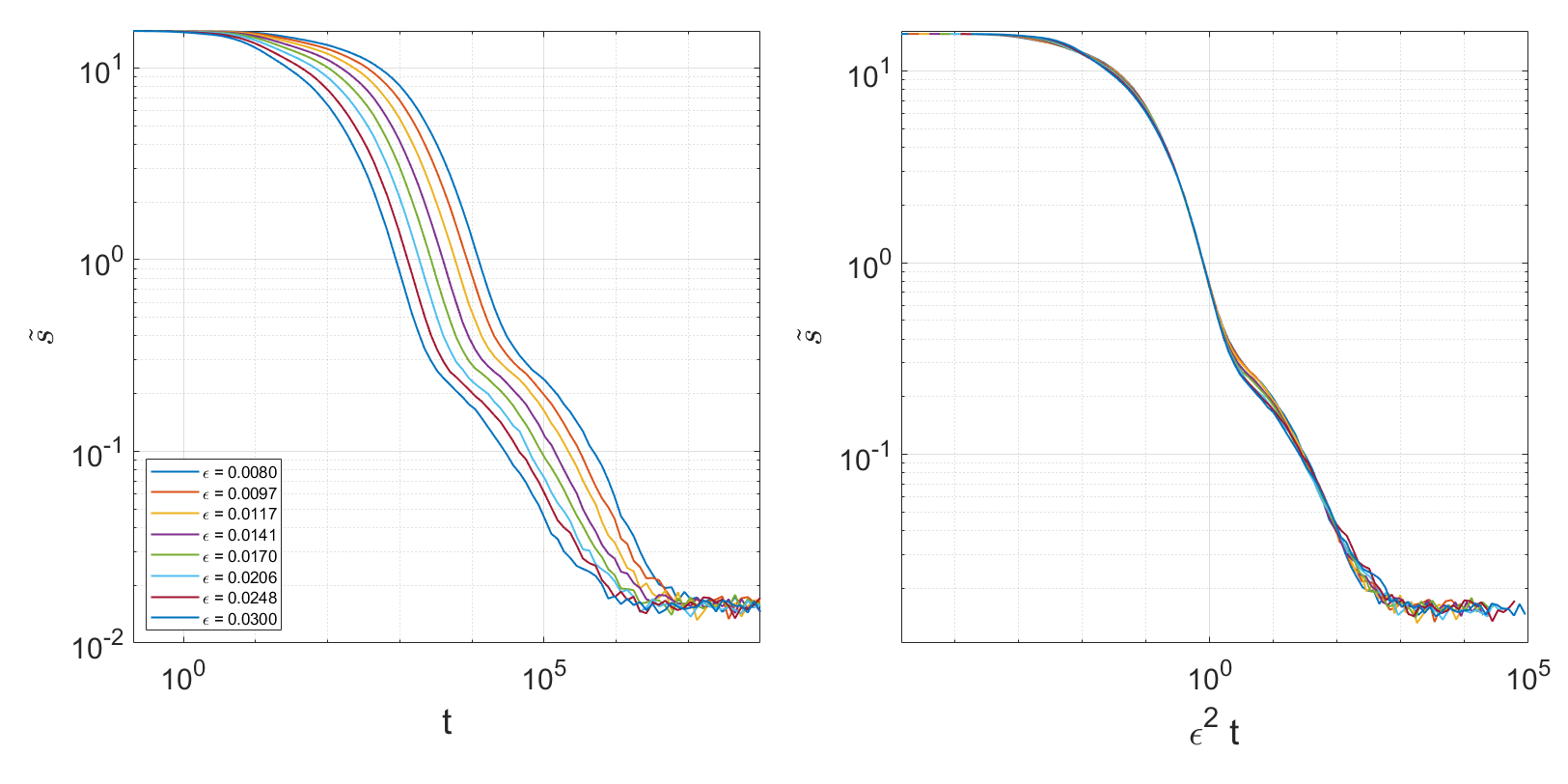}
    \caption{Entropy deficit $\tilde{s}(t) = s_{\text{eq}} - s(t)$ plotted against the unscaled time $t$ (left) and the rescaled time $\epsilon^2 t$ (right) for values of $\epsilon \in [0.0080, 0.0300]$. The collapse of the curves in the right panel confirms the $1/\epsilon^2$ dynamical scaling predicted by the theory.}
    \label{fig:collasso}
\end{figure}

To study the relaxation timescales, we monitor the evolution of the spectral entropy, which in a discrete setting is defined as
\begin{equation*}
    s(t) = \frac{2\pi}{N}\sum_k \log(n_k)\,.
\end{equation*}
According to the $H$-theorem, this quantity is monotonically increasing and saturates at thermal equilibrium. The numerical results confirm our theoretical predictions. As shown in Fig.~\ref{fig:entropy_comparison}, the system with site-dependent coefficients exhibits significantly faster thermalization compared to the standard $\alpha$-FPUT model with constant coefficients ($\alpha_j=1, \chi_j=1$). Furthermore, we observe a dynamical scaling governed by the $\epsilon^2$ factor predicted by the modified wave kinetic equation~\eqref{eq:wavekin}. Indeed, Fig.~\ref{fig:collasso} demonstrates that the entropy deficit $\tilde{s}(t) = s_{\text{eq}} - s(t)$, computed for different values of $\epsilon \in [0.0080, 0.0300]$, collapses onto a single universal curve when plotted against the rescaled time $\epsilon^2 t$.

\section{Conclusion}\label{sec:conclusions}
In this work, we have studied a one-dimensional anharmonic chain. The model is a variant of the $\alpha$-FPUT model, where we let the spring coefficients depend on space, namely on the site position.
We have analyzed the problem in the framework of wave turbulence. Considering that the nonlinearity is small, we can proceed to a perturbative approach.
The goal is to obtain the corresponding wave kinetic equation.
To do that, we first find the resonant manifold for our problem, which turns out to be different from standard $\alpha$-FPUT. Indeed, since the spatial dependence of the coupling coefficient introduces an additional mode associated with the spatial modulation, resonances arise at a lower order than in the standard $\alpha$-FPUT.
The wave kinetic equation is analytically derived and two terms appear at leading order.
A linear term, which corresponds to Bragg scattering, leads to a symmetric spectrum for asymmetric initial conditions. Yet, in the long-time limit this does not contribute.
The nonlinear term is characterized by 3-wave+1  interactions, and has a characteristic timescale proportional to $1/\epsilon^2$, i.e., much faster than the typical time scale for the standard $\alpha$-FPUT which is $1/\epsilon^4$ in the large-box limit \cite{pistone2019universal}. This work can be easily extended to other chains with variable coefficients as the $\beta$-FPUT or the Toda Lattice.

\mo{It is well-known that waves propagating through random media can undergo Anderson localization, whereby interference among multiple scattering paths produces eigenfunctions that are exponentially localized in space. In our framework, we predict strong wave reflection and eventually trapping associated with the Bragg-scattering mechanism. Although these effects are qualitatively reminiscent of localization, our present kinetic description does not retain all the phase correlations required to establish Anderson localization. Therefore, the precise relationship between the wave kinetic approach and Anderson localization remains an open question. Clarifying this connection, possibly through the analysis of spatial correlations and coherent multiple-scattering effects, will be an important subject of future research. }

 \FloatBarrier
\begin{appendices}
\section{Derivation of the wave kinetic equation} \label{app:derivation}

In this Appendix, we formally derive the kinetic equation associated with the dynamical equation \eqref{starting}. 

We consider the model written in Fourier space:
\begin{equation}
\begin{split}
    i\frac{db_1}{dt} &= \omega_1 b_1 + \epsilon \sum_{k_2,k_3} F_{1,2}\chi_3 b_2\delta_{1}^{2,3} \\
    &\quad + \epsilon\sum_{k_2, k_3, k_4}\alpha_4\left( V_{1,2,3}b_2 b_3\delta_1^{2,3,4} + 2 V_{1,-2,3}b_2^*b_3\delta_{1,2}^{3,4}\right).
\end{split}
\end{equation}
First, we introduce the interaction-representation variable $b_1 = c_1 e^{-i\omega_1 t}$ to remove the fast linear oscillations, yielding:
\begin{equation*}
\begin{split}
    i\frac{dc_1}{dt} &= \epsilon \sum_{k_2,k_3} F_{1,2}\chi_3 c_2 e^{-i(\omega_2-\omega_1)t}\delta_{1}^{2,3} \\
    &\quad + \epsilon \sum_{k_2,k_3,k_4} \alpha_4 \left( V_{1,2,3} c_2 c_3 e^{-i(\omega_2+\omega_3-\omega_1)t} \delta_1^{2,3,4} + 2 V_{1,-2,3} c_2^* c_3 e^{-i(\omega_3-\omega_2-\omega_1)t} \delta_{1,2}^{3,4} \right).
\end{split}
\end{equation*}

Next, we compute the time evolution of the mean squared modulus of $c_1$:
\begin{equation} \label{eq:App}
\begin{split}
    \frac{d \langle|c_1|^2\rangle}{dt} &= 2\epsilon \mathrm{Im}\Bigg[\sum_{k_2,k_3} F_{1,2}\chi_3 \langle c_1^*c_2\rangle e^{-i(\omega_2-\omega_1)t} \delta_{1}^{2,3} \\
    &\quad + \sum_{k_2,k_3,k_4} \alpha_4 \left( V_{1,2,3} \langle c_1^* c_2 c_3 \rangle e^{-i\omega_{2,3}^1 t} \delta_1^{2,3,4} + 2 V_{1,-2,3} \langle c_1^* c_2^* c_3 \rangle e^{-i\omega_3^{1,2} t} \delta_{1,2}^{3,4} \right)\Bigg].
\end{split}
\end{equation}

\mo{As with the notation used for the Kronecker deltas, here and in the sequel we adopt a compact notation for frequency combinations, meaning $\omega_{i,j,\dots}^{k,l,\dots} = \omega_i + \omega_j + \dots - \omega_k - \omega_l - \dots$. For instance, $\omega_{2,3}^1 = \omega_2 + \omega_3 - \omega_1$ and $\omega_3^{1,2} = \omega_3 - \omega_1 - \omega_2$.}

At leading order, $d\langle |c_1|^2\rangle/dt = 0$. This is because, under the homogeneity assumption, $\langle c_1^* c_2\rangle = J_1 \delta_1^2$, which makes the first term purely real, while the three-point correlators vanish due to Wick's theorem. Therefore, we must proceed to the next order and compute the time evolution of the correlators:
\begin{align*}
    \frac{d\langle c_1^*c_2\rangle}{dt} &= i\epsilon \sum_{k_4,k_5} F_{1,4}^*\chi_5^* \langle c_4^*c_2\rangle e^{i(\omega_4-\omega_1)t}\delta^{4,5}_1 - i\epsilon \sum_{k_4,k_5} F_{2,4}\chi_5\langle c_1^*c_4\rangle e^{-i(\omega_4-\omega_2)t}\delta_{4,5}^2 \\
    &\quad + i\epsilon\sum_{k_4,k_5,k_6}\alpha_6\left(V_{1,4,5}^* \langle c_4^*c_5^*c_2\rangle e^{i\omega^1_{4,5}t}\delta_1^{4,5,6} + 2V^*_{1,-4,5}\langle c_4c_5^*c_2\rangle e^{i\omega_{5}^{1,4}t}\delta_{1,4}^{5,6}\right) \\
    &\quad - i\epsilon\sum_{k_4,k_5,k_6}\alpha_6\left(V_{2,4,5} \langle c_4c_5c_1^*\rangle e^{-i\omega^1_{4,5}t}\delta_1^{4,5,6} + 2V_{2,-4,5}\langle c_4^*c_5c_1^*\rangle e^{-i\omega_{5}^{1,4}t}\delta_{1,4}^{5,6}\right).
\end{align*}

We note that the combination of the linear and nonlinear terms gives rise to mixed three-point correlators. It will be shown that these terms yield higher-order contributions; therefore, we can safely neglect them.
\begin{equation*}
    \frac{d\langle c_1^*c_2 \rangle }{dt} = i\epsilon \sum_{k_4,k_5} F_{1,4}^*\chi_5^* \langle c_2^*c_4\rangle e^{i(\omega_4-\omega_1)t}\delta_{4,5}^1 - i\epsilon \sum_{k_4,k_5} F_{2,4}\chi_5\langle c_1^*c_4\rangle e^{-i(\omega_4-\omega_2)t}\delta_{4,5}^2 +\mathcal{O}(\epsilon^2).
\end{equation*}

A straightforward time integration from $0$ to $t$ yields:
\begin{align}\label{firstcorr}
    \langle c_1^*c_2\rangle &= \epsilon \sum_{k_5}\left(\frac{e^{i(\omega_2-\omega_1)t}-1}{\omega_2-\omega_1}F_{1,2}^*\chi_5^* J_2 \delta_{2,5}^1 - \frac{e^{i(\omega_2-\omega_1)t}-1}{\omega_2-\omega_1} F_{2,1}\chi_5^* J_1 \delta_{1}^{2,5}\right)\nonumber \\
    &= \epsilon \frac{e^{i(\omega_2-\omega_1)t}-1}{\omega_2-\omega_1}\sum_{k_5}\delta_{1}^{2,5}\chi_5^*\left(F_{1,2}^* J_2 - F_{2,1} J_1 \right).
\end{align}
Here, we have evaluated the time integral assuming the initial condition $\langle c_1^*c_2\rangle(0) = 0$ for $k_1 \neq k_2$. Note that the diagonal case $k_1 = k_2$ yields a strictly real initial condition $\langle |c_1|^2\rangle(0) = J_1$. When substituted back into Eq. \eqref{eq:App}, this diagonal contribution vanishes identically upon taking the imaginary part. Furthermore, we have used the property $\chi^*_k = \chi_{-k}$, which holds since $\chi$ is the Fourier transform of a real-valued function.
Next, we compute the time evolution of the three-point correlator:
\begin{align*}
    \frac{d}{dt} \langle c_1^* c_2 c_3 \rangle &= \left\langle \frac{d c_1^*}{dt} c_2 c_3 \right\rangle + \left\langle c_1^* \frac{dc_2}{d t} c_3 \right\rangle + \left\langle c_1^* c_2 \frac{d c_3}{d t} \right\rangle \\
    &= i\epsilon\sum_{k_4,k_5} \Big( F_{1,4}^* \chi_5^*\langle c_4^*c_2c_3 \rangle e^{-i(\omega_4-\omega_1)t}\delta_1^{4,5} - F_{2,4} \chi_5\langle c_4c_1^*c_3 \rangle e^{i(\omega_4-\omega_2)t}\delta_2^{4,5} \\
    &\quad - F_{3,4} \chi_5\langle c_4c_2c_1^* \rangle e^{i(\omega_4-\omega_3)t}\delta_3^{4,5}\Big) \\
    &\quad + i\epsilon \sum_{k_5,k_6,k_7} \Big( \alpha_7^* V_{1,5,6}^* \langle c_5^* c_6^* c_2 c_3 \rangle e^{i\omega_{5,6}^1 t} \delta_1^{5,6,7} + 2\alpha_7^* V_{1,-5,6}^* \langle c_5 c_6^* c_2 c_3 \rangle e^{i\omega_6^{1,5} t} \delta_{1,5}^{6,7} \\
    &\quad - \alpha_7 V_{2,5,6} \langle c_1^* c_5 c_6 c_3 \rangle e^{-i\omega_{5,6}^2 t} \delta_2^{5,6,7} - 2\alpha_7 V_{2,-5,6} \langle c_1^* c_5^* c_6 c_3 \rangle e^{-i\omega_6^{2,5} t} \delta_{2,5}^{6,7} \\
    &\quad - \alpha_7 V_{3,5,6} \langle c_1^* c_2 c_5 c_6 \rangle e^{-i\omega_{5,6}^3 t} \delta_3^{5,6,7} - 2\alpha_7 V_{3,-5,6} \langle c_1^* c_2 c_5^* c_6 \rangle e^{-i\omega_6^{3,5} t} \delta_{3,5}^{6,7} \Big).
\end{align*}
To evaluate the right-hand side of the equation, we must close the hierarchy of correlators by truncating our perturbative expansion. Specifically, since the right-hand side is already proportional to $\epsilon$, incorporating any time evolution of the correlators would introduce higher-order $\mathcal{O}(\epsilon^2)$ terms. By requiring that their time evolution be zero at this order, we can evaluate them using their initial state at $t=0$. Under the assumption of Random Phase and Amplitudes (RPA), Wick's theorem applies. Consequently, all correlators with an unequal number of $c$ and $c^*$ fields, including the mixed three-point correlators generated by the linear terms and unbalanced fourth-order correlators, vanish identically. Obtaining non-zero contributions from these unbalanced terms would require computing their explicit time evolution, which introduces higher-order corrections in $\epsilon$ that we neglect. We are thus left solely with the standard, fourth-order correlators. 
\begin{align*}
    \langle c_5^* c_6^* c_2 c_3 \rangle &= J_2 J_3 (\delta_5^2 \delta_6^3 + \delta_5^3 \delta_6^2), \\
    \langle c_1^* c_5^* c_6 c_3 \rangle &= J_1 J_3 \delta_1^6 \delta_3^5 + J_1 J_5 \delta_1^3 \delta_5^6, \\
    \langle c_1^* c_2 c_5^* c_6 \rangle &= J_1 J_2 \delta_1^6 \delta_2^5 + J_1 J_5 \delta_1^2 \delta_5^6.
\end{align*}
Substituting these expressions back into the derivative, we find:
\begin{align*}
    \frac{d}{dt} \langle c_1^* c_2 c_3 \rangle &= 2i\epsilon \sum_{k_7} \alpha_7 \left( V_{1,2,3}^* J_2 J_3 \delta_{1,7}^{2,3} - V_{2,-3,1} J_1 J_3 \delta_{2,3}^{1,7} - V_{3,-2,1} J_1 J_2 \delta_{2,3}^{1,7} \right) e^{i\omega_{2,3}^1 t} \\
    &\quad - 2i\epsilon \sum_{k_5}\left(\alpha_2 V_{2,-5,5} J_1 J_5 \delta_1^3 e^{i\omega_2 t} + \alpha_3 V_{3,-5,5} J_1 J_5 \delta_1^2 e^{i\omega_3 t} \right).
\end{align*}
Assuming the initial condition $\langle c_1^* c_2 c_3 \rangle(0) = 0$, we can perform the time integral:
\begin{align}\label{secondcorr}
    \langle c_1^* c_2 c_3 \rangle (t) &= 2\epsilon \sum_{k_7} \alpha_7 \left( V_{1,2,3}^* J_2 J_3 \delta_{1,7}^{2,3} - V_{2,-3,1} J_1 J_3 \delta_{2,3}^{1,7} - V_{3,-2,1} J_1 J_2 \delta_{2,3}^{1,7} \right)\frac{e^{i\omega_{2,3}^1 t} - 1}{\omega_{2,3}^1} \nonumber\\
    &\quad - 2\epsilon \sum_{k_5} \left( \alpha_2 V_{2,-5,5} J_1 J_5 \delta_1^3 \frac{e^{i\omega_2 t} - 1}{\omega_2} + \alpha_3 V_{3,-5,5} J_1 J_5 \delta_1^2 \frac{e^{i\omega_3 t} - 1}{\omega_3} \right).
\end{align}
We apply exactly the same procedure to the other three-point correlator appearing in the equation for the modulus squared, $\langle c_1^* c_2^* c_3 \rangle$. Its time derivative is given by:
\begin{align*}
    \frac{d}{dt} \langle c_1^* c_2^* c_3 \rangle &= \left\langle \frac{d c_1^*}{d t} c_2^* c_3 \right\rangle + \left\langle c_1^* \frac{d c_2^*}{d t} c_3 \right\rangle + \left\langle c_1^* c_2^* \frac{d c_3}{dt} \right\rangle \\
    &= i\epsilon\sum_{k_4,k_5} \Big( F_{1,4}^* \chi_5^*\langle c_4^*c_2^*c_3 \rangle e^{-i(\omega_4-\omega_1)t}\delta_1^{4,5} + F_{2,4}^*\chi_5^*\langle c_4^*c_1c_3^* \rangle e^{-i(\omega_4-\omega_2)t}\delta_2^{4,5} \\
    &\quad - F_{3,4} \chi_5\langle c_4c_2^*c_1^* \rangle e^{i(\omega_4-\omega_3)t}\delta_3^{4,5}\Big)\\
    &\quad + i\epsilon \sum_{k_5,k_6,k_7} \Big( \alpha_7^* V_{1,5,6}^* \langle c_5^* c_6^* c_2^* c_3 \rangle e^{i\omega_{5,6}^1 t} \delta_1^{5,6,7} + 2\alpha_7^* V_{1,-5,6}^* \langle c_5 c_6^* c_2^* c_3 \rangle e^{i\omega_6^{1,5} t} \delta_{1,5}^{6,7} \\
    &\quad + \alpha_7^* V_{2,5,6}^* \langle c_1^* c_5^* c_6^* c_3 \rangle e^{i\omega_{5,6}^2 t} \delta_2^{5,6,7} + 2\alpha_7^* V_{2,-5,6}^* \langle c_1^* c_5 c_6^* c_3 \rangle e^{i\omega_6^{2,5} t} \delta_{2,5}^{6,7} \\
    &\quad - \alpha_7 V_{3,5,6} \langle c_1^* c_2^* c_5 c_6 \rangle e^{-i\omega_{5,6}^3 t} \delta_3^{5,6,7} - 2\alpha_7 V_{3,-5,6} \langle c_1^* c_2^* c_5^* c_6 \rangle e^{-i\omega_6^{3,5} t} \delta_{3,5}^{6,7} \Big).
\end{align*}

We apply Wick's theorem to the three surviving terms:
\begin{align*}
    \langle c_5 c_6^* c_2^* c_3 \rangle &= J_2 J_3 \delta_2^5 \delta_3^6 + J_2 J_5 \delta_2^3 \delta_5^6, \\
    \langle c_1^* c_5 c_6^* c_3 \rangle &= J_1 J_3 \delta_1^5 \delta_3^6 + J_1 J_5 \delta_1^3 \delta_5^6, \\
    \langle c_1^* c_2^* c_5 c_6 \rangle &= J_1 J_2 (\delta_1^5 \delta_2^6 + \delta_1^6 \delta_2^5).
\end{align*}

Substituting these expressions back into the derivative and exploiting the symmetry under the exchange of dummy indices $5 \leftrightarrow 6$, we collect all terms sharing the common phase $\omega_3^{12}$:
\begin{align*}
    \frac{d}{dt} \langle c_1^* c_2^* c_3 \rangle &= 2i\epsilon \sum_{k_7} \alpha_7 \left( V_{1,-2,3}^* J_2 J_3 \delta_{1,2}^{3,7} + V_{2,-1,3}^* J_1 J_3 \delta_{1,2}^{3,7} - V_{3,1,2} J_1 J_2 \delta_{1,2}^{3,7} \right) e^{i\omega_3^{1,2} t} \\
    &\quad + 2i\epsilon \sum_{k_5} \left( \alpha_1^*V_{1,-5,5}^* J_2 J_5 \delta_2^3 e^{-i\omega_1 t} + \alpha_2^*V_{2,-5,5}^* J_1 J_5 \delta_1^3 e^{-i\omega_2 t} \right).
\end{align*}

The time integration yields:
\begin{align}\label{thirdcorr}
    \langle c_1^* c_2^* c_3 \rangle (t) &= 2\epsilon \sum_{k_7} \alpha_7 \left( V_{1,-2,3}^* J_2 J_3 \delta_{1,2}^{3,7} + V_{2,-1,3}^* J_1 J_3 \delta_2^{1,3,7} - V_{3,1,2} J_1 J_2 \delta_{1,2}^{3,7} \right) \frac{e^{i\omega_3^{1,2} t} - 1}{\omega_3^{1,2}}\nonumber \\
    &\quad - 2\epsilon \sum_{k_5} \left( \alpha_1^*V_{1,-5,5}^* J_2 J_5 \delta_3^2 \frac{e^{-i\omega_1 t} - 1}{\omega_1} + \alpha_2^*V_{2,-5,5}^* J_1 J_5 \delta_1^3 \frac{e^{-i\omega_2 t} - 1}{\omega_2} \right).
\end{align}

Substituting Eqs.~\eqref{firstcorr}, \eqref{secondcorr}, and \eqref{thirdcorr} into Eq.~\eqref{eq:App}, we obtain:
\begin{align}\label{kinetic_eq_pre_limit}
    \partial_t J_1 &= 2\epsilon^2 \text{Im} \Bigg\{ \sum_{2,3}|F_{1,2}|^2|\chi_3|^2(J_2-J_1)\frac{1-e^{-i\omega_2^1t}}{\omega_2^1}\delta_1^{2,3}\nonumber\\
    &\quad + 2\sum_{2,3,4} |\alpha_4 V_{1,2,3}|^2 (J_2 J_3 - J_1 J_3 - J_1 J_2) \frac{e^{i\omega_1^{2,3}t} - 1}{\omega_1^{2,3}} \delta_1^{2,3,4} \nonumber\\
    &\quad - 4\sum_{2,3,4} |\alpha_4 V_{1,-2,3}|^2 (J_1 J_2 - J_2 J_3 - J_1 J_3) \frac{e^{i\omega_{1,2}^3 t} - 1}{\omega_{1,2}^3} \delta_{1,2}^{3,4} \nonumber\\
    &\quad + 2\sum_{2,3} \Big( |\alpha_1V_{1,-3,3}|^2 J_2 J_3 \frac{e^{i\omega_1 t} - 1}{\omega_1} \Big) \Bigg\}.
\end{align}

Here we used the symmetries $V^*_{1,2,3} = V_{2,-3,1} = V_{3,-2,1}$, $V^*_{1,-2,3} = V^*_{2,-1,3} = V_{3,1,2}$, and $\alpha_k^* = \alpha_{-k}$ to extract the squared modulus of the coefficients. Taking the large-box limit ($L\rightarrow \infty$), the sums become integrals, and we introduce the continuous wave action density $n(k,t) = L J_k / (2\pi)$.
\mo{Taking the same limit for the coefficients and relabeling the variables, we get:
\begin{equation*}
    |\alpha_k|^2\rightarrow \frac{L}{2\pi}|\alpha_k|^2, \quad    |\chi_k|^2\rightarrow \frac{L}{2\pi}|\chi_k|^2.
\end{equation*}}

Next, we consider the weak-nonlinearity limit ($\epsilon \rightarrow 0$). Defining the slowtime scale $\tau = \epsilon^2 t$, we perform the weak limit:
\begin{equation*}
    \lim_{\epsilon\to 0}\mo{\mathrm{ \,Im}}\left[ \int \frac{e^{i\omega_1^{2,3}\frac{\tau}{\epsilon^2}} - 1}{\omega_1^{2,3}} dk_2dk_3\right] = \int \pi\delta(\Delta\omega_1^{2,3})dk_2dk_3,
\end{equation*}
and similarly for the other expressions.

\begin{align}
    \partial_t n_1 &= 2\pi \int dk_{2,3}|F_{1,2}|^2|\chi_3|^2(n_2-n_1)\delta(\Delta \omega_1^2)\delta_1^{2,3} \nonumber \\
    &\quad + 4\pi \int dk_{2,3,4}\Bigg( |\alpha_4 V_{1,2,3}|^2 (n_2 n_3 - n_1 n_3 - n_1 n_2) \delta(\Delta\omega_1^{2,3})\delta_1^{2,3,4} \nonumber\\
    &\quad - 2|\alpha_4 V_{1,-2,3}|^2 (n_1 n_2 - n_2 n_3 - n_1 n_3) \delta(\Delta\omega_{1,2}^3)\delta_{1,2}^{3,4}\Bigg) \nonumber \\
    &\quad + 4\pi\int dk_{2,3} |\alpha_1V_{1,-3,3}|^2 n_2 n_3 \delta(\Delta\omega_1).
\end{align}

Since $V_{0, -3, 3} = 0$, the last integral vanishes identically. Thus, the modified wave kinetic equation consists of a linear part, accounting for Bragg scattering, and a nonlinear part that incorporates the non-constant coefficient introduced in our model:
\begin{align}
    \partial_t n_1 &= 2\pi \int dk_{2,3}F_{1,2}|\chi_3|^2(F_{1,2}^*n_2-F_{2,1}n_1)\delta(\Delta \omega_1^2)\delta_1^{2,3} \nonumber \\
    &\quad + 4\pi \int dk_{2,3,4}\Bigg( |\alpha_4 V_{1,2,3}|^2 (n_2 n_3 - n_1 n_3 - n_1 n_2) \delta(\Delta\omega_1^{2,3})\delta_1^{2,3,4} \nonumber\\
    &\quad - 2|\alpha_4 V_{1,-2,3}|^2(n_1 n_2 - n_2 n_3 - n_1 n_3) \delta(\Delta\omega_{1,2}^3)\delta_{1,2}^{3,4}\Bigg).
\end{align}

We can rewrite this in a more symmetric form:
\begin{align}\label{simmform}
    \partial_t n_1 &= 2\pi \int dk_{2,3}|F_{1,2}|^2|\chi_3|^2(n_2-n_1)\delta(\Delta \omega_1^2)\delta_1^{2,3} \nonumber \\
    &\quad + 4\pi \int dk_{2,3,4} \Big[ \mathcal{R}^1_{2,3,4} - \mathcal{R}^2_{1,3,4} - \mathcal{R}^3_{1,2,4} \Big],
\end{align}
where 
\begin{equation*}
    \mathcal{R}^1_{2,3,4} = |\alpha_{4}|^2 |V_{1,2,3}|^2 (n_2 n_3 - n_1 n_3 - n_1 n_2)\delta(\Delta\omega_1^{2,3}) \delta_{1}^{2,3,4}.
\end{equation*}

Focusing solely on the linear term, it is instructive to write its explicit form after integrating out the Dirac deltas. From the frequency resonance condition $\Delta\omega_1^2 = 0$, we obtain $\omega_1 = \omega_2$, which corresponds to $k_1=\pm k_2$.
\begin{align*}
    &2\pi \int dk_{2,3}|F_{1,2}|^2|\chi_3|^2(n_2-n_1)\delta(\Delta \omega_1^2)\delta_1^{2,3} \\
    &= 2\pi \int dk_3 \frac{|F_{1,-1}|^2}{|d\omega_{k_1}/dk_1|}|\chi_3|^2 (n_{-1}-n_1) \delta(2k_1-k_3) \\
    &= 2\pi\frac{|F_{1,-1}|^2}{|d\omega_{k_1}/dk_1|}|\chi_{2k_1}|^2 (n_{-1}-n_1).
\end{align*}

Here, we have also integrated out the second delta function. Only the branch $k_1 = -k_2$ gives a non-zero contribution.

\end{appendices}

\backmatter



\bmhead{Acknowledgements}
M.O. was funded by Progetti di Ricerca di Interesse Nazionale (PRIN), Projects No. 2020X4T57A and No. 2022WKRYNL, project ''Mathematical Methods in Non-Linear Physics'' and ''FieldTurb'', Istituto Nazionale di Fisica Nucleare. The Simons Foundation, United States, Award No. 652354 on Wave Turbulence is acknowledged for support. GD is funded by the European Union (ERC, Project Acronym: OPPIWaM, Project No.:
101222068).

\bmhead{Data Availability} No data were created for this submission.

\bmhead{Conflict of interest} The authors report no conflict of interest.




\end{document}